\documentclass[aps,prl,twocolumn,showpacs,
amsmath,amssymb,floatfix]{revtex4} 
 
\newcommand{\beq}{\begin{equation}}   
\newcommand{\eeq}{\end{equation}}   
\newcommand{\bea}{\begin{eqnarray}}   
\newcommand{\eea}{\end{eqnarray}}   
   
\begin{document}      
\title{The twistor geometry of three-qubit entanglement} 
\author{P\'eter L\'evay} 
\affiliation{Department of Theoretical Physics, Institute of Physics, Budapest University of Technology and Economics}  
\date{\today} 
\begin{abstract}
A geometrical description of three qubit entanglement
is given. A part of the transformations corresponding to  stochastic local operations and classical communication on the qubits
is regarded as a gauge degree of freedom.
Entangled states can be represented by the points of the Klein quadric ${\cal Q}$ a space known from twistor theory.  
It is shown that three-qubit invariants are vanishing on special subspaces of ${\cal Q}$.
An invariant vanishing for the $GHZ$ class is proposed.
A geometric interpretation of the canonical decomposition and the inequality
for distributed entanglement is also given. 
\end{abstract}      
\maketitle{} 
 
The basic role played by geometry in such areas of modern physics as general relativity, quantum field theory, and string theory is well-known.
Further examples for the usefulness of geometric ideas for a deeper understanding of quantum theory are provided by the rich research field of geometric phases, fractional spin and statistics \cite{shapere} and holonomic quantum computation \cite{zanardi}. Recently some authors \cite{brody,bengtsson,mosseri,bernevig,miyake,levay} have also tried to understand quantum entanglement in geometric terms. 
In particular it was observed\cite{mosseri,bernevig} that two and a special class of three-qubit entangled states can be described by certain maps that are entanglement sensitive. These maps enable a geometric description of entanglement in terms of fiber bundles. Fiber bundles are spaces that are locally look like the product of two spaces the base and the fiber, globally however, they can exhibit a nontrivial twisted structure. In this picture this twisting of the bundle accounts for some portion of quantum entanglement.
For two qubits using the correspondence between fibre bundles and the language of
gauge fields these ideas were elaborated\cite{levay}. 
The essence of this approach  was to provide a description of entanglement by regarding a part of the LOCC transformations of local operations and classical communication corresponding to a {\it fixed} subsystem as a gauge degree of freedom. 
For example for two qubits our Hilbert space is
${\bf C}^2\otimes{\bf C}^2\simeq {\bf C}^4$ hence the space of normalized states is the seven sphere $S^7$ that can be regarded as a fiber bundle over the four sphere $S^4$. The fiber is $SU(2)\simeq S^3$ which is precisely the group of LOCC
 (gauge) transformations corresponding to one of the subsytems. 
Then two-qubit entanglement as seen from the viewpoint of this subsystem can be described by characterizing submanifolds
of the four sphere $S^4$.

In this letter we generalize this gauge theory motivated approach to understand three-qubit entanglement of three parties A, B and C in geometric terms.
In order to make our presentation more transparent it is
convenient to enlarge our gauge group corresponding to party A from LOCC to SLOCC (i.e. stochastic LOCC) transformations.
As it is well-known two entangled states are SLOCC equivalent if they can be
converted back and forth with (maybe different) nonvanishing probabilities\cite{bennett, vidal}.
In this case the states in question can preform the same tasks in quantum information processing although with different probabilities of success.
Group theoretically this means that if we wish to account for SLOCC equivalence we have to enlarge our group of gauge transformations from $SU(2)$ to $SL(2, {\bf C})$.
Moreover, it is more comfortable to work with unnormalized states hence we do not fix the determinant of the invertible operator corresponding to SLOOC transformations of subsystem A to unity. 
This introduces an extra complex constant, hence in the following we use as the group of gauge transformations the full group of invertible $2\times 2$ complex matrices i.e. $GL(2, {\bf C})$. 

Consider now a three-qubit state $\vert \psi\rangle\in {\bf C}^2\otimes {\bf C}^2\otimes {\bf C}^2$  of the form
\beq
\label{state}
\vert\psi\rangle =\sum_{a,b,c=0,1}C_{abc}\vert abc\rangle,\quad \vert abc\rangle\equiv \vert a\rangle_A\otimes\vert b\rangle_B\otimes\vert c\rangle_C.
\eeq
\noindent
Let us define two four component vectors $Z^{\mu}$ and $W^{\nu}$ $\mu,\nu=1,2,3,4$ as
\beq
\label{zw}
C_{0bc}\equiv \frac{1}{\sqrt{2}} Z^{\mu}E_{\mu bc},\quad
C_{1bc}\equiv \frac{1}{\sqrt{2}} W^{\mu}E_{\mu bc},
\eeq
\noindent
where $E_k=-i{\sigma}_k, k=1,2,3$ and $E_4=I$ with $I$ the $2\times 2$ unit matrix and ${\sigma}_k $ are the Pauli matrices. Notice that the vectors $Z$ and $W$ are the components of $C_0$ and $C_1$ in a basis equivalent to the magic base of Hill and Wootters\cite{wootters}.
If these vectors are linearly independent they define a complex two-plane in ${\bf C}^4$. 
We define two scalar products on ${\bf C}^4$ as $\langle Z\vert W\rangle\equiv
\overline{Z}^{\mu}W_{\mu}$ and $Z\cdot W\equiv Z^{\mu}W_{\mu}$ (summation for repeated indices is understood). We also use the notation $\vert\vert Z\vert\vert^2=\langle Z\vert Z\rangle$ etc.
Calculating the reduced density matrix ${\rho}_A={\rm Tr}_{BC}\vert\psi\rangle\langle\psi\vert$ in terms of $Z$ and $W$ shows that ${\rm Det}{\rho}_A=
\vert\vert Z\vert\vert^2\vert\vert W\vert\vert^2-\vert\langle Z\vert W\rangle\vert^2$. Since ${\rm Det}{\rho}_A=0$ iff $\vert\psi\rangle$ is $A(BC)$ separable 
, by virtue of the Cauchy-Schwartz inequality we see that this can only happen
if $Z=\lambda W$ i.e. iff $Z$ and $W$ are linearly dependent.
Hence excluding this case (which is trivial from the point of view of party A)
we are led to reformulating three-qubit entanglement by studying the structure of the manifold of two-planes in ${\bf C}^4$, i.e. the four complex dimensional Grassmannian $Gr(2, 4)$.

A group element $g\in GL(2, {\bf C})$ with matrix elements $\alpha,\beta,\gamma,\delta$ acts on the pair of vectors $Z$ and $W$ as
$Z^{\mu}\mapsto {\alpha}Z^{\mu}+{\beta}W^{\mu}$, $W^{\mu}\mapsto {\gamma}Z^{\mu}+{\delta}W^{\mu}$. It is easy to show that the action of $GL(2, {\bf C})$ on the pair of linearly independent vectors $(Z,W)$ is free and transitive hence we can regard the space of such pairs $(Z,W)$ as a fiber bundle  
over $Gr(2,4)$ with gauge group $GL(2, {\bf C})$.
Since we would like to understand the geometry of three-qubit entanglement
by examining the structure of subspaces of $Gr(2,4)$ it is convenient to introduce coordinates for this space. Let us define these coordinates (Pl\"ucker coordinates) as
\beq
\label{plucker}
P^{\mu\nu}\equiv Z^{\mu}W^{\nu}-Z^{\nu}W^{\mu}
\eeq
\noindent
i.e. $P^{\mu\nu}$ is a {\it separable} bivector.
These $6$ complex coordinates transform under the gauge group $GL(2, {\bf C})$ as
$P^{\mu\nu}\mapsto (\alpha\delta-\beta\gamma)P^{\mu\nu}$, hence they are defined only up to a complex scalar factor. Notice, however that $P$ is invariant with respect to the $SL(2, {\bf C})$ subgroup of SLOCC transformations. Moreover, since the Pl\"ucker relation
\beq
\label{pluckerrel}
P^{12}P^{34}-P^{13}P^{24}+P^{23}P^{14}=0 
\eeq
\noindent
holds, we can remove all the redundancies accounting for the $4$ complex dimensions of $Gr(2,4)$. 
Note that (\ref{pluckerrel}) defines a quadric ${\cal Q}$ (the Klein-quadric) embedded in ${\bf C}P^5$. It can be shown\cite{ward} that (\ref{pluckerrel}) is a sufficient and necessary condition for an arbitrary bivector $P$ to be separable i.e. of the (\ref{plucker}) form. Hence we can also look at $Gr(2,4)$ as the Klein-quadric ${\cal Q}$
in ${\bf C}P^5$ which is the space encoding the information we need on the geometry of three-qubit entanglement.

In order to extract this information we will study special values of both LOCC and SLOCC invariants.
Let us first consider the SLOCC invariant three-tangle\cite{ckw}
${\tau}_{ABC}\equiv 4\vert D(C)\vert$ where
\begin{eqnarray}
 D(C)&\equiv &  C_{000}^2C_{111}^2+C_{001}^2C_{110}^2+C_{010}^2C_{101}^2+C_{011}^2C_{100}^2\nonumber\\
 &-&2(C_{000}C_{001}C_{110}C_{111}+C_{000}C_{010}C_{101}C_{111}\nonumber\\&+&
 C_{000}C_{011}C_{100}C_{111}
 +C_{001}C_{010}C_{101}C_{110}\nonumber\\&+&
 C_{001}C_{011}C_{110}C_{100}+C_{010}C_{011}C_{101}C_{100})\nonumber\\
 &4&(C_{000}C_{011}C_{101}C_{110}+C_{001}C_{010}C_{100}C_{111}).
 \label{hypdet}
 \end{eqnarray}
\noindent
is the Cayley hyperdeterminant\cite{gelfand}. Due to the method of Schl\"afli\cite{gelfand}
we can express $D(C)$ as the discriminant of the quadratic form ${\rm Det}(xC_0+yC_1)$ in the variables $x$ and $y$ as
$D(C)=({\rm Tr}C_0{\rm Tr}C_1-{\rm Tr}(C_0C_1))^2-4{\rm Det}(C_0){\rm Det}(C_1)$.
For a complex $2\times 2$ matrix $M$ defining its "adjoint" as $M^{\prime}\equiv {\rm Det}(M)M^{-1}$ this expression can be written as $D(C)=[{\rm Tr}(C^{\prime}_0C_1)]^2-4{\rm Det}(C^{\prime}_0C_1)$. 
Noting that $E_4^{\prime} =E_4$, $E_k^{\prime}=-E_k$ and using (\ref{zw}) we obtain the following nice expression for the three tangle
\beq
\label{3tangle}
{\tau}_{ABC}=2\vert P^{\mu\nu}P_{\mu\nu}\vert=4\vert (Z\cdot Z)(W\cdot W)-(Z\cdot W)^2\vert.
\eeq
\noindent
Using (\ref{3tangle}) the $SL(2, {\bf C})\times SL(2, {\bf C})\times SL(2, {\bf C})\simeq SL(2, {\bf C})\times SO(4, {\bf C})$ invariance of ${\tau}_{ABC}$ can be immediately established. Moreover, we have also learnt that the residual SLOCC transformations on parties B and C are represented by the adjoint action of $SO(4, {\bf C})$ on the separable bivector $P\in{\cal Q}$ in the form $P\mapsto SPS^T, S\in SO(4, {\bf C})$. 

In order to gain further insight into the geometry of three-qubit entanglement
we write out $4{\rm Det}{\rho}_A$ which is called\cite{ckw} the "concurrence between qubit A and the pair BC" in terms of $P$. Denoting this quantity by ${\tau}_{A(BC)}$ we get
\beq
\label{conca}
{\tau}_{A(BC)}=2P^{\mu\nu}\overline{P}_{\mu\nu}=4(\vert\vert Z\vert\vert^2\vert\vert W\vert\vert^2-\vert\langle Z\vert W\rangle\vert^2).
\eeq
\noindent
Expressions (\ref{3tangle}) and (\ref{conca}) show that both quantities ${\tau}_{ABC}$ and ${\tau}_{A(BC)}$ are invariant with respect to $SL(2, {\bf C})$ SLOCC transformations performed by party A and they pick up a constant under $GL(2, {\bf C})$ transformations. Since (\ref{pluckerrel}) is invariant with respect to multiplication by scalar factors ${\tau}_{ABC}$ and ${\tau}_{A(BC)}$ define real valued
functions on the Klein quadric ${\cal Q}$.
It is also amusing to see that in this geometric setting the Coffman-Kundu-Wootters inequality of distributed entanglement\cite{ckw} can be related to the triangle inequality in Pl\"ucker coordinates.
Indeed, from (\ref{3tangle}) and (\ref{conca}) we immediately see that ${\tau}_{ABC}\leq {\tau}_{A(BC)}$, i.e. the residual tangle is smaller or equal than the amount of entanglement of qubit $A$ with the pair $BC$. 

Let us calculate now the quantities ${\tau}_-={\tau}_{B(AC)}\equiv 4{\rm Det}{\rho}_B$ and ${\tau}_+={\tau}_{C(AB)}\equiv 4{\rm Det}{\rho}_C$ in terms of $Z$ and $W$. Straightforward calculation shows
\beq
\label{rhobc}
{\tau}_{\pm}=\vert Z\cdot Z\vert^2+\vert W\cdot W\vert^2+2\vert Z\cdot W\vert^2+(P^{\mu\nu}\mp \ast P^{\mu\nu})\overline{P}_{\mu\nu},
\eeq
\noindent
where
$\ast P$ is the dual of the bivector $P$ i.e. $\ast P_{\mu\nu}\equiv \frac{1}{2}
{\varepsilon}_{\mu\nu\varrho\sigma}P^{\varrho\sigma}$.
Note, that these quantities unlike the previous ones are not invariant under $SL(2, {\bf C})$ transformations of party $A$, however their vanishing is still a gauge invariant notion.

Let us look now on the reduced density matrix ${\rho}_{BC}={\rm Tr}_A\vert\psi\rangle\langle \psi\vert$! It is easy to show that in the magic base it takes the form ${\rho}_{BC}^{\mu\nu}=Z^{\mu}\overline{Z^{\nu}}+W^{\mu}\overline{W^{\nu}}$.
Introducing the spin-flipped operator\cite{ckw} as $\tilde{\rho}\equiv (\sigma_2\otimes \sigma_2)\overline{\rho}(\sigma_2\otimes\sigma_2)$ which in the magic base 
amounts to merely complex conjugation, we obtain the formula
\beq
\label{rhobesc}
{\rm Tr}( {\rho}_{BC}\tilde{\rho}_{BC})=\vert Z\cdot Z\vert^2+\vert W\cdot W\vert^2+2\vert Z\cdot W\vert^2.
\eeq
\noindent
From Eqs. (\ref{rhobc}) and (\ref{rhobesc}) we see that ${\rm Tr}({\rho}_{BC}\tilde{\rho}_{BC})=2({\rm Det}{\rho}_B+{\rm Det}{\rho}_C-{\rm Det}{\rho}_A)$ as it has to be\cite{ckw}.
After a cyclic permutation of this relation we see that
\beq
\label{masikketto}
{\rm Tr}({\rho}_{\pm}\tilde{\rho}_{\pm})=(P^{\mu\nu}\pm \ast P^{\mu\nu})\overline{P}_{\mu\nu},
\eeq
\noindent
where
$ {\rho}_+\equiv{\rho}_{AB}$ and 
${\rho}_-\equiv{\rho}_{AC}.$
Eqs. (\ref{masikketto}) are expressible entirely in terms of the bivector $P$  
showing that they are gauge invariant up to a factor.
Notice that the squared difference of the square-root of the two nonzero eigenvalues of ${\rho}_{\pm}\tilde{\rho}_{\pm}$
define the two-tangles ${\tau}_{AB}$ and ${\tau}_{AC}$
subject to the Coffman-Kundu-Wootters relation\cite{ckw} ${\tau}_{A(BC)}={\tau}_{ABC}+{\tau}_{AB}+{\tau}_{AC}$. All the quantities in this relation
are gauge invariant up to a complex factor hence this relation
is characterizing relationships between special submanifolds of the Klein quadric ${\cal Q}$. What are these special submanifolds? 

Consider first an arbitrary separable bivector $P$ giving rise to the plane
$\lambda Z^{\mu}+\kappa W^{\mu}$. Then we can find the principal null directions of this plane by solving the quadratic equation ${\lambda}^2(Z\cdot Z)^2+2{\lambda\kappa}(Z\cdot W)+{\kappa}^2(W\cdot W)^2$. According to (\ref{3tangle}) the discriminant of this equation is just the Cayley hyperdeterminant so we have two principal directions for ${\tau}_{ABC}\neq 0$ and one  for ${\tau}_{ABC}=0$.
One can show that these directions are given by the formula $U^{\mu}_{\pm}=Z_{\nu}P^{\nu\mu}\pm\frac{1}{2}\sqrt{{\tau}_{ABC}}e^{i\varphi/2}Z^{\mu}$, where
$\varphi\equiv\arg[{(Z\cdot W)^2-(Z\cdot Z)(W\cdot W)}]$. It is obvious that
for normalized states finding the canonical form of a three-qubit state
in the method of Acin et.al.\cite{acin} is equivalent to rotating one of the vectors $Z$ or $W$ to one of these principal directions using the LOCC subgroup of
our gauge group then applying further $LOCC$ transformations. This observation also
accounts for the two possibilities for this decomposition (apart from the case when ${\tau}_{ABC}=0$ when this decomposition is unique). Moreover, the complex phase phactor appearing in the canonical form\cite{acin}
can be related to our $\varphi$ as defined above. 

Let us now characterize geometrically submanifolds of ${\cal Q}$ correspondingto $B(AC)$ and $C(AB)$ separable states.
For such states we have ${\tau}_-$ or ${\tau}_+$ equals zero.
Let us chose a subset of two-planes characterized by the conditions
$\ast P=-P$ and $Z\cdot W=0$, i.e. the bivector $P$ is anti-self-dual and is defined by two orthogonal vectors.
Notice that such vectors are automatically null i.e. $Z\cdot Z=W\cdot W=0$ as can be seen by contracting the equation of anti-self-duality with $Z$ and $W$. 
Subspaces of this form are called ${\beta}$-planes in twistor theory\cite{sen}.
Hence from (\ref{rhobc}) we see that for $\beta$-planes ${\tau}_-=0$.
However, we can even state more after noticing that the four summands in (\ref{rhobc}) are all nonnegative. Indeed taking into account the nonegativity of the left hand side of (\ref{masikketto}) a moment thought shows that ${\tau}_-=0$ {\it precisely} for $\beta$-planes. A similar line of reasoning shows that ${\tau}_+=0$
precisely for $\alpha$-planes characterized by the analogous conditions and $P=\ast P$ i.e. self duality.
Let us illustrate this on the $B(AC)$ separable state $\vert\psi_-\rangle=
\alpha\vert 100\rangle +\gamma\vert 001\rangle$. In this case $Z^{\mu}=\frac{\gamma}{\sqrt{2}}(i,-1,0,0)^T$ and $W^{\nu}=\frac{\alpha}{\sqrt{2}}(0,0,i,1)^T$.
These vectors are null, orthogonal and a calcualtion shows $\ast P=-P$. 
Similarly for the $C(AB)$ separable state $\vert \psi_+\rangle =\alpha\vert 100\rangle +\beta\vert 010\rangle$ we have 
$Z^{\mu}=\frac{\beta}{\sqrt{2}}(i,1,0,0)^T$ and $W^{\nu}$ is just the same as above. Here the change of sign in $Z$ results in the condition $\ast P=P$.
Clearly all $\alpha$ and $\beta$ planes through the origin of ${\bf C}^4$ can be obtained as $SO(4,{\bf C})$ orbits of a "canonical" $P$ corresponding to either of the states $\vert\psi_{\pm}\rangle$. Since two-planes in ${\bf C}^4$ are represented by points on ${\cal Q}$, $SO(4, {\bf C})$ orbits of the form $SPS^T$ then give rise to submanifolds of ${\cal Q}$ after factoring out with those transformations which fix the canonical $P$.   
One can show\cite{sen} that the set of ${\alpha}$ planes in ${\cal Q}$ can be parametrized by the three-dimensional complex projective space ${\bf C}P^3$. Similarly the set of $\beta$-planes is another ${\bf C}P^3_{\ast}$ which is the projective dual of the previous one\cite{sen}.

Consider next the Werner state $\vert W\rangle =\alpha\vert 100\rangle+\beta\vert 010\rangle +\gamma\vert 001\rangle$.
For this state we have $Z^{\mu}=\frac{1}{\sqrt{2}}((\beta+\gamma)i,\beta-\gamma,0,0)^T$, $W^{\nu}=\frac{\alpha}{\sqrt{2}}(0,0,i,1)^T$.
We see that $Z$ is not null but orthogonal to the null vector $W$.
Moreover, the null vector $W$ lies in the intersection of the $\alpha$ and $\beta$ planes of the previous paragraph. The separable bivector $P$ corresponding to $\vert W\rangle$ has the important property ${\tau}_{ABC}=0$ as one can quickly check from (\ref{3tangle}).
Separable bivectors satisfying ${\tau}_{ABC}=0$ are called \textit{null separable bivectors} or null-twistors in twistor theory.
Let us show that the aforementioned properties characterize
precisely the Werner-class! 
For this define the quadratic form $Q(P_1,P_2)\equiv{\varepsilon}_{\mu\nu\varrho\sigma}P^{\mu\nu}_1P^{\varrho\sigma}_2$. Clearly relation (\ref{pluckerrel}) is equivalent to $Q(P,P)=0$, moreover one has ${\tau}_{ABC}=2\vert Q(P,\ast P)\vert$. Then one can easily prove the lemma that the intersection of two planes $P_1$ and $P_2$ is nonzero if and only if $Q(P_1, P_2)=0$.
From this lemma it follows that ${\tau}_{ABC}=0$ precisely when the plane $P$
intersects with its dual plane $\ast P$. Moreover for $\alpha$ and $\beta$-planes by virtue of $\ast P=\pm P$, ${\tau}_{ABC}=0$.  
Now the Werner class is characterized by the conditions\cite{vidal} ${\tau}_{A(BC)}\neq 0$, ${\tau}_{\pm}\neq 0$ and ${\tau}_{ABC}=0$. The first three of this conditions excludes the possibilities of collinear $Z$ and $W$ and $\alpha$ and $\beta$-planes.
The planes $P$ and $\ast P$ corresponding to these cases are either degenerate or beeing identical up to sign  hence intersect trivially. 
Then we see that the Werner class is characterized precisely by the condition
that $P$ and $\ast P$ \textit{ intersect along a line}. This was what we illustrated for the state $\vert W\rangle$. 
Notice that equations $Q(P,P)=Q(\ast P, \ast P)$ hold trivially, showing that $P$ and $\ast P$ are points lying on the Klein quadric ${\cal Q}$.
Condition $Q(P, \ast P)=0$ characterizing the Werner class shows that 
the tangent vector $\ast P$ of the tangent plane of ${\cal Q}$ at $P$ lies entirely in ${\cal Q}$. Moreover, the line $ L\equiv tP+(1-t)\ast P$  through $P$ and $\ast P$ also satisfying $Q(L,L)=0$ lies entirely in the Klein quadric. Such lines are called in twistor theory \textit{null lines}\cite{ward}.
Now ${\cal Q}$ can be regarded as the compactification and complexification of
Minkowski spacetime. From twistor theory it is well-known that the null lines of ${\cal Q}$ represent the light-cone (conformal) structure of Minkowski spacetime. Hence we found an interesting connection between three-qubit states belonging to the Werner class and special null lines in  Minkowski space time.
Is there any deeper physical reason for this connection?

How to characterize geometrically the $GHZ$ class, which is known to be the other class from the two inequivalent ones\cite{vidal} characterizing genuine three-qubit entanglement?
In order to attack this problem we recall that there is still one more independent invariant we have not discussed. This is the Kempe
invariant characterizing hidden nonlocalities (i.e. ones that cannot be revealed by inspection of local density matrices)\cite{kempe}.
This invariant\cite{kempe,sudbery} rewritten in a form convenient for our purposes is
\beq
\label{kempeinv}
\xi_{ABC}={\rm Tr}({\cal A}^3+{\cal B}^3+3{\cal C}^{\dagger}{\cal C}{\cal A}
+3{\cal C}{\cal C}^{\dagger}{\cal B}),
\eeq
\noindent
where ${\cal A}\equiv C_0C^{\dagger}_0$, ${\cal B}\equiv C_1C^{\dagger}_1$
and ${\cal C}\equiv C_0C^{\dagger}_1$.
After a straightforward but tedious calcuation we get
$\xi_{ABC}={\cal N}^3+\frac{3}{8}\left({\omega}_{ABC}-\Lambda_{ABC}\right)$,
where ${\cal N}=\langle\psi\vert\psi\rangle$,
$\Lambda_{ABC}\equiv
{\cal N}({\tau}_{A(BC)}+{\tau}_{B(AC)}+{\tau}_{C(AB)})$, and
$\omega_{ABC}=4{\rm Tr}({\rho}_{BC}\overline{P}P)$.
Now we claim that the new permutation and LOCC invariant characterizing the $GHZ$ class is
\beq
\label{sigma}
\sigma_{ABC}\equiv {\cal N}{\tau}_{ABC}-4{\rm Tr}(\rho_{BC}\overline{P}P).
\eeq
\noindent
In order to prove this notice that $\sigma_{ABC}$ can be written in terms of the principal null directions as $\frac{1}{2}(\vert\vert U_+\vert\vert^2+\vert\vert U_-\vert\vert^2+\vert\vert V_+\vert\vert^2+\vert\vert V_-\vert\vert^2)\equiv\vert\vert U\vert\vert^2+\vert\vert V\vert\vert^2$. Here we observe that the principal null directions $V_{\pm}^{\mu}\equiv P^{\mu\nu}W_\nu\pm\frac{1}{2}\sqrt{{\tau}_{ABC}}e^{{i\varphi}/2}W^{\mu}$ are up to a complex number the same as the corresponding ones  $U^{\mu}_{\mp}$ defined previously. 
${\sigma}_{ABC}$ is LOCC and permutation invariant and nonegative.
It is zero iff $U=V=0$ i.e. when $Z$ and $W$ are precisely the \textit{two different} principal null directions. Such vectors beeing the eigenvectors of $P^{\mu\nu}$ with the \textit{nonzero eigenvalues} $\pm\frac{1}{2}{\sqrt{{\tau}_{ABC}}}e^{i\varphi/2}$ are clearly null
hence $Z\cdot Z=W\cdot W =0$. Moreover, the eigenvalues are nonzero hence $Z\cdot W\neq 0$.
The conditions characterizing the $GHZ$ class\cite{vidal} are ${\tau}_{A(BC)}\neq 0$, ${\tau}_{\pm}\neq 0$, and ${\tau}_{ABC}\neq 0$. 
These conditions can be shown to follow from ${\sigma}_{ABC}\equiv 0$,
hence the vanishing of this invariant characterizes the $GHZ$ class.
In order to check these statements, take the standard $GHZ$ state $\vert GHZ\rangle=\alpha\vert 000\rangle +\beta\vert 111\rangle$. For this state we have
$Z^{\mu}=\frac{\alpha}{\sqrt{2}}(0,0,i,1)^T$ and $W^{\nu}=\frac{\beta}{\sqrt{2}}(0,0,-i,1)^T$. These vectors are null, $Z\cdot W\neq 0$ and ${\tau}_{ABC}\neq 0$ 
, they are the null directions of $P$ with the only nonvanishing component $P^{34}=i\alpha\beta$, and an explicit calculation shows that ${\sigma}_{ABC}=0$. 
It is well-known\cite{acin,gingrich} that for normalized states
${\tau}_{A(BC)}$, ${\tau}_{\pm}$ and ${\tau}_{ABC}$ are entanglement monotones satisfying the inequalities
$0\leq {\tau}_{A(BC)},{\tau}_{\pm},{\tau}_{ABC}\leq 1$. Using $\vert{\rm Tr}(A^{\dagger}B)\vert\leq ({\rm Tr}(A^{\dagger}A))^{1/2}({\rm Tr}(B^{\dagger}B)^{1/2}$ and ${\rm Tr}(\overline{P}P)^2=2({\rm Det}{\rho}_A)^2$ one can prove that $0\leq {\sigma}_{ABC}\leq 1$ as well.
We conjecture that ${\sigma}_{ABC}$ is the missing one from the list of entanglement monotones \cite{gingrich}.

In this paper we regarded a $GL(2, {\bf C})$ part of SLOCC acting on qubit $A$ as a gauge degree of freedom. This naturally leaded us to the manifold of two-planes in ${\bf C}^4$ or the Klein-quadric.
We have found that some of the LOCC invariants (${\tau}_{ABC}$, ${\tau}_{A(BC)}$, ${\tau}_{AB}$, $\tau_{AC}$) are gauge invariants up to a constant hence
can be defined as real-valued functions on ${\cal Q}$, others (${\tau}_{\pm}$) though do not exhibit a gauge-invariant meaning, their vanishing however, is still a gauge invariant notion. Armed with this observation we managed to describe different SLOCC classes of entanglement in geometric terms. 
On geometric grounds we have proposed a new invariant characterizing the GHZ class. We illuminated the geometric meaning of the canonical decomposition, and the inequality for distributed entanglement.
We remark that our method can be generalized for $n$-qubits. In this case we have to consider the Grassmannian $Gr(2, 2^{n-1})$ of two planes in ${\bf C}^{2^{n-1}}$ as the base of a $GL(2, {\bf C})$ fiber bundle. It is also interesting to note that even LOCC orbits can be described in this framework using a bundle over
$Gr(2, 2^{n-1})$ with a four sphere as the fiber\cite{battaglia}. 
Twistor methods were originally introduced to physics to describe the causal structure of space-time.
It is of fundamental importance to understand why these methods seem to be tailor made to describe n-qubit entanglement too. 

Financial support from the Orsz\'agos Tudom\'anyos Kutat\'asi Alap (OTKA), (grant numbers T032453 and T038191) is gratefully acknowledged.


\begin{thebibliography}{} 
\bibitem{shapere} A. Shapere and F. Wilczek (eds.) {\it Geometric Phases in Physics} Wiley and Sons (1989).
\bibitem{zanardi} P. Zanardi and M. Rasetti, Phys. Lett. \textbf{A264}, 94 (1999).
\bibitem{brody} D. C. Brody and L. P. Hughston, J. Geom. Phys. \textbf{38}, 19 (2001).
\bibitem{bengtsson} I. Bengtsson, J. Br\"annlund and K. Zyczkowsi, Int. J. Mod. Phys. \textbf{A17}, 4675 (2002).
\bibitem{mosseri} R. Mosseri and R. Dandoloff, J. Phys. \textbf{A34}, 10243 (2001).
\bibitem{bernevig} B. A. Bernevig and H. D. Chen, J. Phys. \textbf{A36}, 8325
(2003).
\bibitem{miyake} A. Miyake, Phys. Rev. \textbf{A67}, 012108 (2003).
\bibitem{levay} P. L\'evay, J. Phys. \textbf{A37} 1821 (2004).
\bibitem{bennett} C. H. Bennett \textit{et al.}, Phys. Rev. \textbf{A63}, 012307 (2000).
\bibitem{vidal} W. D\"ur, G. Vidal, and J. I. Cirac, Phys. Rev. \textbf{A62}, 062314 (2000).
\bibitem{wootters} S. Hill and W. Wootters, Phys. Rev. Lett. \textbf{78}, 5022 (1997).
\bibitem{ward} R. S. Ward, R. O. Wells jr., \textit{Twistor geometry and field theory}, Cambridge monographs on mathematical physics (1990).
\bibitem{ckw} V. Coffman, J Kundu and W. K. Wootters, Phys. Rev. \textbf{A61}
052306 (2000).
\bibitem{gelfand} I. M. Gelfand, M. M. Kapranov, A. V. Zelevinsky, \textit{Discriminants, resultants and multidimensional determinants}, Birkh\"auser Boston 1994.
\bibitem{sen} C. Nash, S. Sen,\textit{Topology and geometry for physicists}, Academic Press 1983. 
\bibitem{acin} A. Acin, A. Andrianov, L. Costa, E. Jan\'e, J. I. Latorre and R. Tarrach, Phys. Rev. Lett. \textbf{85}, 1560 (2000).
\bibitem{kempe} J. Kempe, Phys. Rev. \textbf{A60}, 910 (1999).
\bibitem{sudbery} T. Sudbery, J. Phys. \textbf{A34}, 643 (2001).
\bibitem{gingrich} R. M. Gingrich, Phys. Rev. \textbf{A65}, 052302 (2002).
\bibitem{battaglia} F. Battaglia, Proc. Am. Math. Soc. \textbf{124}, 2185 (1996).
\end{thebibliography}
\end{document}